# Cortical Xi-Alpha model for resting state electric neuronal activity

Roberto D. Pascual-Marqui[1], Kieko Kochi[1], Toshihiko Kinoshita[2]

1: The KEY Institute for Brain-Mind Research; Department of Psychiatry, Psychotherapy, and Psychosomatics; University of Zurich, Switzerland
2: Department of Neuropsychiatry, Kansai Medical University, Osaka, Japan

**Corresponding author: RD Pascual-Marqui**
robertod.pascual-marqui@uzh.ch ; https://www.uzh.ch/keyinst
https://scholar.google.com/citations?user=DDqjOkUAAAAJ

## 1. Abstract

The power spectra of awake resting state EEG recordings in humans typically have an Alpha peak at around 10 Hz riding a decreasing background "Xi process". Normal and pathological variations occur in the form of more than one peak or none at all. The single channel Xi-Alpha model (Pascual-Marqui et al 1988, https://doi.org/10.3109/00207458808985730) separated these two processes and provided an adequate low-dimensional parametric description of EEG spectra. Currently lacking is a generative whole cortex model for activity spectra and intracortical functional connectivity for each process. Here we introduce the "cortical Xi-Alpha model" for such a purpose. The cross-spectral density matrices are modeled as additive components, where each one consists of a scalar spectrum that multiplies a frequency invariant Hermitian covariance matrix for the cortical functional connectivity structure. This model has very low dimension, and despite its simple "separation of variables" form, it offers a very rich repertoire of diverse spatio-spectral properties, as well as diverse whole cortex functional connectivity patterns. The scalp EEG model conserves the same separation of variables form, allowing simple estimation from scalp to cortex.

Two independent open-access, resting state eyes open and closed EEG data sets (203 participants with 61 electrodes, and 47 participants with 26 electrodes) were used to demonstrate, test, and validate the model. Results summary:
- The average dimension of the vectorized cross-spectra lies between two and three for the whole cortex, justifying two processes as an adequate approximation for resting state EEG.
- A non-negative matrix factorization analysis of population power spectra sampled at 6239 cortical grey matter voxels with only two components (identified as Xi and Alpha) explains 99% of the variance.
- The median value of explained variance was 95% for the "cortical Xi-Alpha model" across all datasets and conditions.
- The Alpha process is more strongly located in posterior cortical regions, while the Xi process is more distributed and leans towards frontal regions.
- The Xi lagged connectivity matrix for cortical sources is isotropic with interdistance, while Alpha is not.
- Laminar recordings suggest that pyramidal neurons in layers 2/3 generate the Xi-process, while pyramidal neurons in layers 5/6 generate the Alpha process.

## 2. Introduction

The power spectra of awake resting state EEG recordings in humans typically have an Alpha peak at around 10 Hz riding a smoothly decreasing background process. Normal and pathological variations occur in the form of more than one peak or none at all (see e.g. Tatum 2021).

A parametric model for EEG spectra was published in 1988 (Pascual-Marqui et al 1988), known as the Xi-Alpha model, where these two processes were represented in additive form:

**Eq. 1**     $\phi(\omega) = \xi(\omega) + \alpha(\omega)$

where $\phi(\omega)$ is the spectral density of a single channel EEG signal at discrete frequency ω, and:

**Eq. 2**     $\xi(\omega) = \dfrac{b}{\left(1 + c\omega^2\right)^d}$

**Eq. 3**     $\alpha(\omega) = \dfrac{e}{\left(1 + f(\omega - \omega_\alpha)^2\right)^g}$

with non-negative parameters "$b,c,d,e,f,g$ and $\omega_\alpha$", where "$\omega_\alpha$" is the alpha peak frequency, "$b,e$" are related to height, "$c,f$" are inversely related to square width, and "$d,g$" are related to power law exponents for each process. These equations constitute an equivalent description of those in Pascual-Marqui et al 1988. This model is defined for $\omega > 0$.

Eq. 2 and Eq. 3 are equivalent in form to the *t*-probability density (except for a scale factor). For large values of "$g$" (e.g. $g \geq 20$), the Alpha peak in Eq. 3 has the form (except for a scale factor) of a Gaussian probability density, i.e.:

**Eq. 4**     $\lim\limits_{g \to \infty}: \alpha(\omega) = e' \exp\left[-f'(\omega - \omega_\alpha)^2\right]$

At high frequencies, with $(g > d)$, the spectrum in Eq. 1 (given Eq. 2 and Eq. 3) obeys a power law in the form:

**Eq. 5**     $\begin{cases} \lim\limits_{\omega \to \infty}: \phi(\omega) = \left(\dfrac{b}{c^d}\right)\left(\dfrac{1}{\omega^{2d}}\right) \\ or \ equivalently: \\ \lim\limits_{\omega \to \infty}: \ln\phi(\omega) = \ln\left(\dfrac{b}{c^d}\right) - 2d \ln\omega \end{cases}$

which has been abundantly described in the literature, see e.g. Miller et al (2009), He et al (2010), and He (2014).

A generalization of the Xi-Alpha model was mentioned in Pascual-Marqui et al (1988) for more peaks, e.g. a beta peak:

**Eq. 6**     $\phi(\omega) = \xi(\omega) + \alpha(\omega) + \beta(\omega)$

and for the absence of peaks:

**Eq. 7**     $\phi(\omega) = \xi(\omega)$

with:

**Eq. 8**     $\beta(\omega) = \dfrac{e_\beta}{\left(1 + f_\beta(\omega - \omega_\beta)^2\right)^{g_\beta}}$

Figure 1 shows the graph for the Xi-Alpha model in Eq. 1 (with Eq. 2 and Eq. 3), for typical parameter values declared in the legend. As a sanity check, a comparison can be made between the Xi-Alpha spectrum and an estimated EEG signal spectrum, as in e.g. Chen et al (2020), Figure 4b therein, corresponding to eyes closed resting state condition.

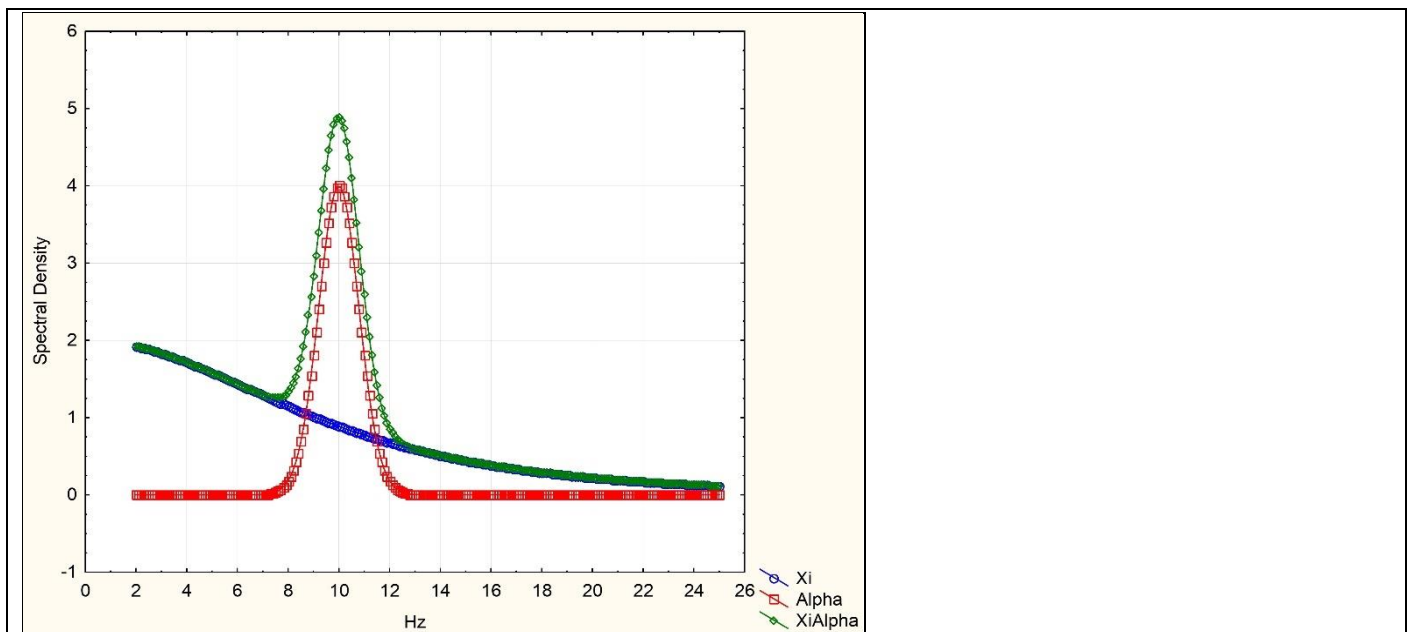

Figure 1: Spectral density $\phi(\omega)$ as a function of discrete frequency $\omega$. Eq. 2 Xi: blue; Eq. 3 Alpha: red; Eq. 1 Xi-Alpha: green. Parameters: b=2; c=0.005; d=2; e=4; f=0.02; g=40; $\omega_\alpha$=10.

Related work in the fitting of spectral peaks and the Xi background process, also known as periodic and aperiodic components, appears in e.g. Whitten et al (2011), Haller et al (2018), Hu and Valdes-Sosa (2019), Donoghue et al (2020), and Lendner et al (2020).

When "fitting" a model to "data", it is common practice to specify a "cost function" that quantitatively measures the deviation between model and data. This is how best fitting parameters are estimated, by minimizing a cost function. In this sense, in Pascual-Marqui et al (1988), maximum likelihood estimation of the Xi-Alpha model in Eq. 1, Eq. 2, and Eq. 3 was used; see e.g. Mardia et al (1979) for likelihood estimation methods, and Brillinger (2001) for the probability distribution of spectra. A similar approach is used in Hu and Valdes-Sosa (2019). In contrast, the algorithms used in Whitten et al (2011), Haller et al (2018), Donoghue et al (2020), and Lendner et al (2020) lack an explicit statement of the cost function for model fitting.

Spectral peak fitting with background Xi (aperiodic) activity is a highly nonlinear problem. Successful estimation is made difficult in the case of two or more overlapping peaks, as studied in Gerster et al (2022). Note that the possible presence of low power white noise can be included in the Xi-Alpha model (Eq. 1) by simply adding a constant corresponding to the white noise spectral power.

The power law exponent "$(2d)$" in Eq. 5 has been hypothesized to be related to excitatory/inhibitory balance at the neuronal circuit level. However, it remains uncertain if the relation is direct or its complete contrary (i.e. inverse). For instance, according to Lombardi et al (2017), an increase in "excitatory/inhibitory" ratio leads to an increase in "$(2d)$". The complete opposite can be found in Gao et al (2017). This controversial issue has been pointed out in (Miskovic et al 2019). Furthermore, it has been demonstrated experimentally in the results of Colombo et al (2019), where induced unresponsiveness leads to an increase or decrease of the power law exponent depending on the anesthetic agent.

In Pascual-Marqui et al (1988), a multichannel EEG Xi-Alpha model was also proposed:

**Eq. 9** $$\Phi(\omega) = [\Xi(\omega)][\mathbf{P}_\xi][\Xi(\omega)] + [\mathbf{A}(\omega)][\mathbf{P}_\alpha][\mathbf{A}(\omega)]$$

where $\Phi(\omega) \in \mathbb{C}^{N_E \times N_E}$ denotes the Hermitian EEG cross-spectral density matrix for $N_E$ electrodes, $\Xi(\omega), \mathbf{A}(\omega) \in \mathbb{R}^{N_E \times N_E}$ are positive definite diagonal matrices with elements:

**Eq. 10** $$\begin{cases}[\Xi(\omega)]_{ii} = \xi_i^{1/2}(\omega) \\ [\mathbf{A}(\omega)]_{ii} = \alpha_i^{1/2}(\omega) \end{cases}, \text{ for } i = 1...N_E$$

where $\xi_i(\omega)$ and $\alpha_i(\omega)$ are the Xi and Alpha spectra at the i-th electrode, and where $\mathbf{P}_\xi, \mathbf{P}_\alpha \in \mathbb{C}^{N_E \times N_E}$ denote the Hermitian coherence matrices for the scalp Xi and Alpha processes which do not change with frequency ω.

Unfortunately, the multichannel EEG Xi-Alpha model in Eq. 9 does not have an equivalent form for cortical grey matter electric neuronal activity, i.e. there is no equivalence in form for a generative cortical Xi-Alpha model.

This study describes a simple generative cortical Xi-Alpha model, whose form is invariant when transformed to the scalp EEG (or extracranial MEG) domain, and which can be estimated using the same complexity of the single signal Xi-Alpha model (i.e. estimation of background plus peaks, i.e. aperiodic plus periodic processes).

Two independent open-access, resting state eyes open and closed EEG data sets (203 participants with 61 electrodes, and 47 participants with 26 electrodes) were used to demonstrate, test, and validate the model. Results show:
- The average dimension of the vectorized cross-spectra lies between two and three for the whole cortex, justifying two processes as an adequate approximation for resting state EEG.
- A non-negative matrix factorization analysis of population power spectra sampled at 6239 cortical grey matter voxels with only two components explains 99% of the variance.
- The median value of explained variance by the new cortical Xi-Alpha model was 95% for the full cross-spectra, across all datasets and conditions.
- The Alpha process is more strongly located in posterior cortical regions, while the Xi process is more distributed and leans towards frontal regions.
- The Xi lagged connectivity matrix for cortical sources is isotropic with interdistance, while Alpha is not.
- Based on recent results from laminar recordings by Mendoza-Halliday et al (2022), it is inferred that pyramidal neurons in layers 2/3 generate the Xi-process, while pyramidal neurons in layers 5/6 generate the Alpha process.

The LORETA-KEY free academic software (https://www.uzh.ch/keyinst/) is used throughout.

## 3. The "cortical Xi-Alpha model"

Let $\mathbf{J}(t) \in \mathbb{R}^{N_V \times 1}$ denote the time varying electric neuronal activity at $N_V$ cortical grey matter voxels, and let $\mathbf{\Psi}(\omega) \in \mathbb{C}^{N_V \times N_V}$ denote its Hermitian cross-spectral density matrix at discrete frequency ω.

The cortical Xi-Alpha model proposed here is written as:

Eq. 11 $\quad \mathbf{\Psi}(\omega) = \xi(\omega)\mathbf{\Psi}_\xi + \alpha(\omega)\mathbf{\Psi}_\alpha$

where $\mathbf{\Psi}_\xi, \mathbf{\Psi}_\alpha \in \mathbb{C}^{N_V \times N_V}$ are the frequency-independent Hermitian covariances for the Xi (ξ) and the Alpha (α) processes, and with $\xi(\omega)$ and $\alpha(\omega)$ given by Eq. 2 and Eq. 3 above.

The model for more than one peak is:

Eq. 12 $\quad \mathbf{\Psi}(\omega) = \xi(\omega)\mathbf{\Psi}_\xi + \alpha(\omega)\mathbf{\Psi}_\alpha + \beta(\omega)\mathbf{\Psi}_\beta + \cdots$

where, e.g.:

Eq. 13 $\quad \beta(\omega) = \dfrac{e_\beta}{\left(1 + f_\beta(\omega - \omega_\beta)^2\right)^{g_\beta}}$

The model for no peaks is:

Eq. 14 $\quad \mathbf{\Psi}(\omega) = \xi(\omega)\mathbf{\Psi}_\xi$

Eq. 12 and Eq. 13 cover the generalization of the cortical Xi-Alpha model.

## 4. Important properties of the cortical Xi-Alpha model

As seen from Eq. 11, the cortical cross-spectral density matrices are modeled as additive components, where each one consists of a scalar spectrum that multiplies a frequency invariant Hermitian covariance matrix for the cortical functional connectivity structure. This model has very low dimension, and despite its simple "separation of variables" form, it offers a very rich repertoire of diverse spatio-spectral properties. As will be shown in the next section, the scalp EEG model conserves the same properties.

Note that there is only one Xi-spectral shape and one Alpha-spectral shape common to the whole cortex. However, each cortical voxel has its unique amplitude for each process, i.e. the amplitudes of each process are not constant over the cortex. Indeed, the amplitudes are given by the diagonal elements of $\mathbf{\Psi}_\xi$ and $\mathbf{\Psi}_\alpha$. Thus, at any selected cortical voxel, the spectrum is a weighted linear combination of the Xi and Alpha spectra, where the frequency for the maximum of the combined Xi and Alpha spectra changes with the weights. This property allows for a rich variety of spectra at each voxel.

Furthermore, the functional connectivity pattern is determined by a weighted sum of two distinct cortical covariance structures (one for each process), which also ensures a very rich connectivity spectra that changes with frequency.

## 5. The corresponding "EEG Xi-Alpha model"

As before, let $\mathbf{J}(t) \in \mathbb{R}^{N_V \times 1}$ denote the time varying electric neuronal activity at $N_V$ cortical grey matter voxels. And now let $\boldsymbol{\phi}(t) \in \mathbb{R}^{N_E \times 1}$ denote the corresponding time varying scalp electric potential differences at $N_E$ electrodes (i.e. EEG), which is determined by the forward EEG equation:

**Eq. 15** $\quad \boldsymbol{\phi}(t) = \mathbf{K}\mathbf{J}(t)$

where $\mathbf{K} \in \mathbb{R}^{N_E \times N_V}$ is the lead field. See e.g. Pascual-Marqui (2007a, 2009).

The "EEG Xi-Alpha model" corresponding to the cortical generative model in Eq. 11 is:

**Eq. 16** $\quad \boldsymbol{\Phi}(\omega) = \xi(\omega)\boldsymbol{\Phi}_\xi + \alpha(\omega)\boldsymbol{\Phi}_\alpha$

with:

**Eq. 17** $\quad \begin{cases} \boldsymbol{\Phi}(\omega) = \mathbf{K}\boldsymbol{\Psi}(\omega)\mathbf{K}^T \\ \boldsymbol{\Phi}_\xi = \mathbf{K}\boldsymbol{\Psi}_\xi\mathbf{K}^T \\ \boldsymbol{\Phi}_\alpha = \mathbf{K}\boldsymbol{\Psi}_\alpha\mathbf{K}^T \end{cases}$

where the superscript "$T$" denotes transpose, $\boldsymbol{\Phi}(\omega) \in \mathbb{C}^{N_E \times N_E}$ denotes the EEG Hermitian cross-spectral density matrix at discrete frequency $\omega$, and $\boldsymbol{\Phi}_\xi, \boldsymbol{\Phi}_\alpha \in \mathbb{C}^{N_E \times N_E}$ denote are the frequency-independent Hermitian EEG covariances for the Xi ($\xi$) and the Alpha ($\alpha$) processes.

Note that $\xi(\omega)$ and $\alpha(\omega)$ are the same for both the cortical (Eq. 11) and the EEG (Eq. 16) Xi-Alpha model.

Note that this also generalizes in a straightforward manner to more than one peak or none at all.

## 6. Estimation outline of the cortical Xi-Alpha model

The first step consists of estimating the corresponding "EEG Xi-Alpha model" in Eq. 16 which uses observable, measurable data, i.e. the EEG cross-spectral matrices. In practice, the EEG cross-spectral matrices can be estimated as the average periodogram over EEG epochs, computed via the discrete Fourier transform using a Hann taper (see e.g. Brillinger (2001), Oppenheim and Schafer (2014), Frei et al (2001), Pascual-Marqui et al (2021)).

The second step consists of using an inverse operator, e.g. eLORETA as in Pascual-Marqui (2007a) and Pascual-Marqui et al (2011), which allows the estimation of the cortical model in Eq. 11.

With respect to the first step, the trace of the corresponding "EEG Xi-Alpha model" in Eq. 16 gives:

**Eq. 18** $\quad u(\omega) = \tau_\xi \xi(\omega) + \tau_\alpha \alpha(\omega) = \dfrac{(\tau_\xi b)}{(1 + c\omega^2)^d} + \dfrac{(\tau_\alpha e)}{(1 + f(\omega - \omega_\alpha)^2)^g}$

with:

**Eq. 19** $\quad \begin{cases} u(\omega) = tr[\boldsymbol{\Phi}(\omega)] \\ \tau_\xi = tr[\boldsymbol{\Phi}_\xi] \\ \tau_\alpha = tr[\boldsymbol{\Phi}_\alpha] \end{cases}$

Note that Eq. 18 is identical in form to the 1988 (Pascual-Marqui et al) single channel Xi-Alpha model (as in Eq. 1, Eq. 2, and Eq. 3), since the traces $\tau_\xi$ and $\tau_\alpha$ can be absorbed into the "*b,e*" parameters in Eq. 2 and Eq. 3. This form allows least squares estimation of the parameters $\left[ (\tau_\xi b), c, d, (\tau_\alpha e), f, g, \omega_\alpha \right]$. For instance, given data $\hat{u}(\omega)$ for the trace of the EEG cross-spectrum, the least squares functional to be minimized with respect to the parameters in the right-hand side of Eq. 18 is:

**Eq. 20** $\quad F = \sum_\omega \left[ \hat{u}(\omega) - u(\omega) \right]^2$

with:

**Eq. 21** $\quad \hat{u}(\omega) = tr\left[ \hat{\boldsymbol{\Phi}}(\omega) \right]$

where $\left[ \hat{\boldsymbol{\Phi}}(\omega) \right]$ denotes e.g. the average EEG multichannel periodogram (see e.g. Brillinger (2001), Oppenheim and Schafer (2014), Frei et al (2001), Pascual-Marqui et al (2021)).

Minimization of the functional in Eq. 20 leads to a highly non-linear set of equations.

After finding the best fit (minimum of Eq. 20), the explained variance (%ExpVar) for the spectrum is defined as:

**Eq. 22** $\quad \%ExpVar_1 = 100\left( 1 - \frac{F_{min}}{F_0} \right)$

where $F_{min}$ is the minimum of Eq. 20 and:

**Eq. 23** $\quad F_0 = \sum_\omega \left[ \hat{u}(\omega) \right]^2$

Let $\hat{\xi}(\omega)$ and $\hat{\alpha}(\omega)$ denote the estimated Xi and Alpha spectra based on the estimated parameters. Note that these estimators are the same for the cortical and for the EEG Xi-Alpha model.

Given the EEG cross-spectrum data $\hat{\boldsymbol{\Phi}}(\omega)$ and the estimated $\left[ \hat{\xi}(\omega), \hat{\alpha}(\omega) \right]$, the following least squares functional is to be minimized with respect to the frequency independent EEG Hermitian covariance matrices $\left[ \boldsymbol{\Phi}_\xi, \boldsymbol{\Phi}_\alpha \right]$:

**Eq. 24** $\quad G = \sum_\omega tr\left\{ \left[ \hat{\boldsymbol{\Phi}}(\omega) - \hat{\xi}(\omega)\boldsymbol{\Phi}_\xi - \hat{\alpha}(\omega)\boldsymbol{\Phi}_\alpha \right]^2 \right\}$

Minimization of the functional in Eq. 24 leads to a set of linear equations. The estimated parameters are denoted $\left[ \hat{\boldsymbol{\Phi}}_\xi, \hat{\boldsymbol{\Phi}}_\alpha \right]$.

After finding the best fit (minimum of Eq. 24), the explained variance (%ExpVar) for the full cross-spectral matrices is defined as:

**Eq. 25** $\quad \%ExpVar_2 = 100\left( 1 - \frac{G_{min}}{G_0} \right)$

where $G_{min}$ is the minimum of Eq. 24 and:

**Eq. 26** $\quad G_0 = \sum_\omega tr\left\{ \left[ \hat{\boldsymbol{\Phi}}(\omega) \right]^2 \right\}$

Finally, the second step estimates cortical Xi-Alpha model in Eq. 11. For this purpose, a pseudo-inverse of the lead field matrix, denoted as $\mathbf{T} \in \mathbb{R}^{N_V \times N_E}$, can be used:

Eq. 27
$$\begin{cases} \hat{\mathbf{\Psi}}_\xi = \mathbf{T} \hat{\mathbf{\Phi}}_\xi \mathbf{T}^T \\ \hat{\mathbf{\Psi}}_\alpha = \mathbf{T} \hat{\mathbf{\Phi}}_\alpha \mathbf{T}^T \end{cases}$$

The choice of inverse in this work is eLORETA, because of its properties in terms of minimal localization error, minimal false positive activity, and minimal false positive connectivity as compared to a wide range of linear inverse solutions, as shown in Pascual-Marqui et al (2018a).

The estimated cortical Xi-Alpha model is written as:

Eq. 28     $\hat{\mathbf{\Psi}}(\omega) = \hat{\xi}(\omega) \hat{\mathbf{\Psi}}_\xi + \hat{\alpha}(\omega) \hat{\mathbf{\Psi}}_\alpha$

Note that the estimation methods used here are based on unambiguous cost functions to be minimized, which should at least theoretically guarantee optimal parameters that match as best as possible the model to the data.

Note that the estimation methods used here can be applied to the more general cases of several peaks or none at all.

Finally, Appendix A contains the detailed technical description of the estimation method, which is necessary for independent validation, testing, checking, and replicating all the results presented in this work.

### 6.A. Cortical activity and intracortical connectivity

The diagonal elements of $\left[ \hat{\mathbf{\Psi}}_\xi, \hat{\mathbf{\Psi}}_\alpha \right]$, i.e. $\left\{ \left[ \hat{\mathbf{\Psi}}_\xi \right]_{vv}, \left[ \hat{\mathbf{\Psi}}_\alpha \right]_{vv} \right\}$ for $v = 1...N_V$ where $N_V$ is the number of cortical voxels, correspond to the cortical activity distribution of the Xi process and of the Alpha process. These processes have cortical activity distributions that do not change with frequency. The spectrum at the v-th cortical voxel, which is a function of frequency, is a linear combination (with positive weights) of the diagonal elements:

Eq. 29     $\left[ \hat{\mathbf{\Psi}}(\omega) \right]_{vv} = \hat{\xi}(\omega) \left[ \hat{\mathbf{\Psi}}_\xi \right]_{vv} + \hat{\alpha}(\omega) \left[ \hat{\mathbf{\Psi}}_\alpha \right]_{vv}$

Let $\left[ \hat{\mathbf{R}}(\omega) \right]$ denote the coherence matrix corresponding to $\left[ \hat{\mathbf{\Psi}}(\omega) \right]$. And let $\left[ \hat{\mathbf{R}}_\xi, \hat{\mathbf{R}}_\alpha \right]$ denote the Hermitian coherence matrices for each process corresponding to $\left[ \hat{\mathbf{\Psi}}_\xi, \hat{\mathbf{\Psi}}_\alpha \right]$. The coherences of the Xi process and of the Alpha process $\left[ \hat{\mathbf{R}}_\xi, \hat{\mathbf{R}}_\alpha \right]$ do not change with frequency.

In the cortical connectivity analyses for $\left[ \hat{\mathbf{R}}_\xi, \hat{\mathbf{R}}_\alpha \right]$ to be addressed below, use is made of the lagged coherence measure (equivalently denoted as lagged connectivity measure). This measure, developed and introduced in Pascual-Marqui (2007b, 2007c), was derived from an appropriate modeling of the zero lag and the lagged contributions to the coherence between two time series. Lagged coherence eliminates the volume conduction artifact in an optimal way as shown by its theoretical and empirical properties derived Hindriks (2021), and by its empirical properties when compared to a large set of other connectivity measures shown in Pascual-Marqui et al (2018b).

## 7. Exploratory method for the discovery of processes based on non-negative matrix factorization (NMF)

The non-negative spectral power values at each cortical voxel $v = 1 \ldots N_V$ and at each discrete $\omega = 1 \ldots N_\Omega$ can be expressed as the matrix $\Upsilon \in \mathbb{R}^{N_V \times N_\Omega}$, with elements:

**Eq. 30** $\quad [\Upsilon]_{v\omega} = [\Psi(\omega)]_{ii} \geq 0$

If the cortical Xi-Alpha model in Eq. 11 holds, then:

**Eq. 31** $\quad \Upsilon = [diagv(\Psi_\xi)]\xi^T + [diagv(\Psi_\alpha)]\alpha^T$

where the operator $diagv(\bullet)$ takes a square matrix and returns a vector formed by its diagonal elements, and where the vectors $\xi, \alpha \in \mathbb{R}^{N_\Omega \times 1}$ have elements:

**Eq. 32** $\quad \begin{cases} [\xi]_\omega = \xi(\omega) \\ [\alpha]_\omega = \alpha(\omega) \end{cases}$

The model in Eq. 31 has the form a non-negative matrix factorization. See e.g. Paatero and Tapper (1994); Lee and Seung (1999); Pascual-Montano et al (2006). Eq. 31 corresponds to two components, but this form generalizes to more processes.

The exploratory method consists of estimating the cortical spectral power values from the EEG cross-spectrum:

**Eq. 33** $\quad \hat{\Psi}(\omega) = \mathbf{T}[\hat{\Phi}(\omega)]\mathbf{T}^T \in \mathbb{C}^{N_V \times N_V}$

From this, the "voxel by frequency matrix" in Eq. 30 is composed and analyzed via NMF. Note that NMF is applied to data, which is not informed of the characteristics of the Xi-Alpha model.

If the actual underlying generation of cortical electric neuronal activity corresponds to the postulated Xi-Alpha model, then, in a completely blind fashion, the expected results for two components are:
1. The spectral parts of NMF (as in Eq. 32) have the shapes of a decreasing spectrum (Xi) and a peaked spectrum (Alpha);
2. The spatial parts of NMF corresponding to $[diagv(\Psi_\xi)]$ and $[diagv(\Psi_\alpha)]$ in Eq. 31 have cortical generators corresponding to those estimated from fitting the Xi-Alpha model to the data (see Eq. 18 through Eq. 28).

## 8. Estimating the number of distinct cortical processes based on Shannon's entropy

Vectorizing the "cortical Xi-Alpha model" in Eq. 11 gives:

**Eq. 34** $\quad \Psi^{vec}(\omega) = \xi(\omega)\Psi_\xi^{vec} + \alpha(\omega)\Psi_\alpha^{vec}$

with $\Psi^{vec}(\omega), \Psi_\xi^{vec}, \Psi_\alpha^{vec} \in \mathbb{C}^{(N_V^2) \times 1}$. This means that the cortical EEG cross-spectrum lies on a linear subspace of dimension two for two processes. For the general Xi-Alpha model with $N_P$ processes (one Xi process plus $(N_P - 1)$ oscillatory processes), the cortical cross-spectra lies on subspace of dimension $N_P$.

Note from Eq. 11 and Eq. 16, both cortical and EEG XI-Alpha models have the same linear subspace dimension. Therefore, for EEG:

**Eq. 35** $\quad \mathbf{\Phi}^{vec}(\omega) = \xi(\omega)\mathbf{\Phi}_\xi^{vec} + \alpha(\omega)\mathbf{\Phi}_\alpha^{vec}$

with $\mathbf{\Phi}^{vec}(\omega), \mathbf{\Phi}_\xi^{vec}, \mathbf{\Phi}_\alpha^{vec} \in \mathbb{C}^{(N_E^2) \times 1}$.

Consider the Hermitian matrix corresponding to the covariance of the estimated full EEG cross-spectral Hermitian matrices:

**Eq. 36** $\quad \hat{\mathbf{M}} = \frac{1}{N_\Omega} \sum_{\omega=1}^{N_\Omega} \left[\hat{\mathbf{\Phi}}^{vec}(\omega)\right]\left[\hat{\mathbf{\Phi}}^{vec}(\omega)\right]^* \in \mathbb{C}^{(N_E^2) \times (N_E^2)}$

where the superscript "*" denotes transpose and complex conjugate. This matrix should have only two non-zero, positive eigenvalues $(\lambda_1, \lambda_2 > 0)$ for the model in Eq. 34 and Eq. 35.

A general estimator for the number of processes (i.e. for the number of components) is based on Shannon's entropy for the eigenvalues $\lambda'_k$ of $\hat{\mathbf{M}}$:

**Eq. 37** $\quad \hat{N}_P = \exp\left(-\sum_{k=1}^{N_E^2} \lambda_k \ln \lambda_k\right)$

where:

**Eq. 38** $\quad \lambda_k = \left(\sum_{i=1}^{N_E^2} \lambda'_i\right)^{-1} \lambda'_k$

See e.g. Wackermann (1996) equation 2 therein; Camiz and Pillar (2018) equation 7 therein; and Cangelosi and Goriely (2007) equation 20 therein.

## 9. EEG data from Babayan et al (2019)

EEG data from Babayan et al (2019) was used in this study. Details about the subjects and the recordings can be found in the original paper. In summary, the resting state EEG recordings consisted of alternating 60 seconds eyes-closed and 60 seconds eyes open conditions, 16 minutes total, in $N_S = 203$ participants, using $N_E = 61$ scalp electrodes. Preprocessing was performed for artifact correction, with the data downsampled from 2500 Hz to 250 Hz, and band-pass filtered 1 Hz to 45 Hz.

The first 180 seconds of eyes-closed (EC) EEG, and the first 180 seconds of eyes open (EO) EEG, were used here, in the form of $N_W = 180$ epochs of one-second duration each, per condition (EC, EO), per participant. Each one-second epoch consisted of $N_T = 250$ time-samples for $N_E = 61$ scalp electrodes.

The average reference EEG cross-spectrum of each participant was estimated in the form of the average multivariate periodogram over 180 epochs, using a Hann taper (see e.g. Brillinger (2001), Oppenheim and Schafer (2014), Frei et al (2001), Pascual-Marqui et al (2021)).

A "power of two" fast discrete Fourier transform algorithm was used. Average reference one-second EEG epochs were centered to zero mean over time (zero DC level), Hann tapered, and zero padded to $N_T = 256$ (in this order). All analyses were conducted in the frequency range 2Hz to 44Hz. Discrete frequencies $\omega = 1...N_\Omega$, with $N_\Omega = 44$, correspond to actual cycles per second (Hz) as follows:

**Eq. 39**  $\omega_{Hz} = (\omega + 1)(250/256)$

## 10. EEG data from van Dijk et al (2022)

EEG data from Dijk et al (2022) was also used in this study. Details about the subjects and the recordings can be found in the original paper. In summary, the resting state EEG recordings consisted of 2 minutes eyes-closed and 2 minutes eyes open conditions, in $N_S = 47$ healthy control participants, using $N_E = 26$ scalp electrodes. Recordings were sampled at 500 Hz, bandpass filtered between 0.5 to 100 Hz, and notch-filtered at 50 Hz. Preprocessing was performed for artifact correction.

Two minutes of eyes-closed (EC) EEG, and two minutes of eyes open (EO) EEG, were used here, in the form of $N_W = 120$ epochs of one-second duration each, per condition (EC, EO), per participant. Each one-second epoch consisted of $N_T = 500$ time-samples for $N_E = 26$ scalp electrodes.

The average reference EEG cross-spectrum of each participant was estimated in the form of the average multivariate periodogram over 120 epochs, using a Hann taper (see e.g. Brillinger (2001), Oppenheim and Schafer (2014), Frei et al (2001), Pascual-Marqui et al (2021)).

A "power of two" fast discrete Fourier transform algorithm was used. Average reference one-second EEG epochs were centered to zero mean over time (zero DC level), Hann tapered, and zero padded to $N_T = 512$ (in this order). All analyses were conducted in the frequency range 2Hz to 44Hz. Discrete frequencies $\omega = 1...N_\Omega$, with $N_\Omega = 44$, correspond to actual cycles per second (Hz) as follows:

**Eq. 40**  $\omega_{Hz} = (\omega + 1)(500/512)$

## 11. Results and Discussion

*11.A. Individual alpha peak frequency (IAF) from the Xi-Alpha model (theoretical analysis)*

The cortical Xi-Alpha model postulates one unique Xi process and one unique alpha process, common to the whole cortex (in general there can be more peaks), i.e. Eq. 2 and Eq. 3 in Eq. 16. However, they combine with different amplitudes (i.e. heights) at different voxels, producing different "individual alpha peak frequencies (IAF)" for the combination. This is illustrated in Figure 2. Note that this can explain, for instance, the observed differences in IAFs for occipital and frontal regions, corresponding to higher amplitude and faster IAF for occipital, with lower amplitude and slower IAF for frontal regions. See e.g. Sauseng et al (2005).

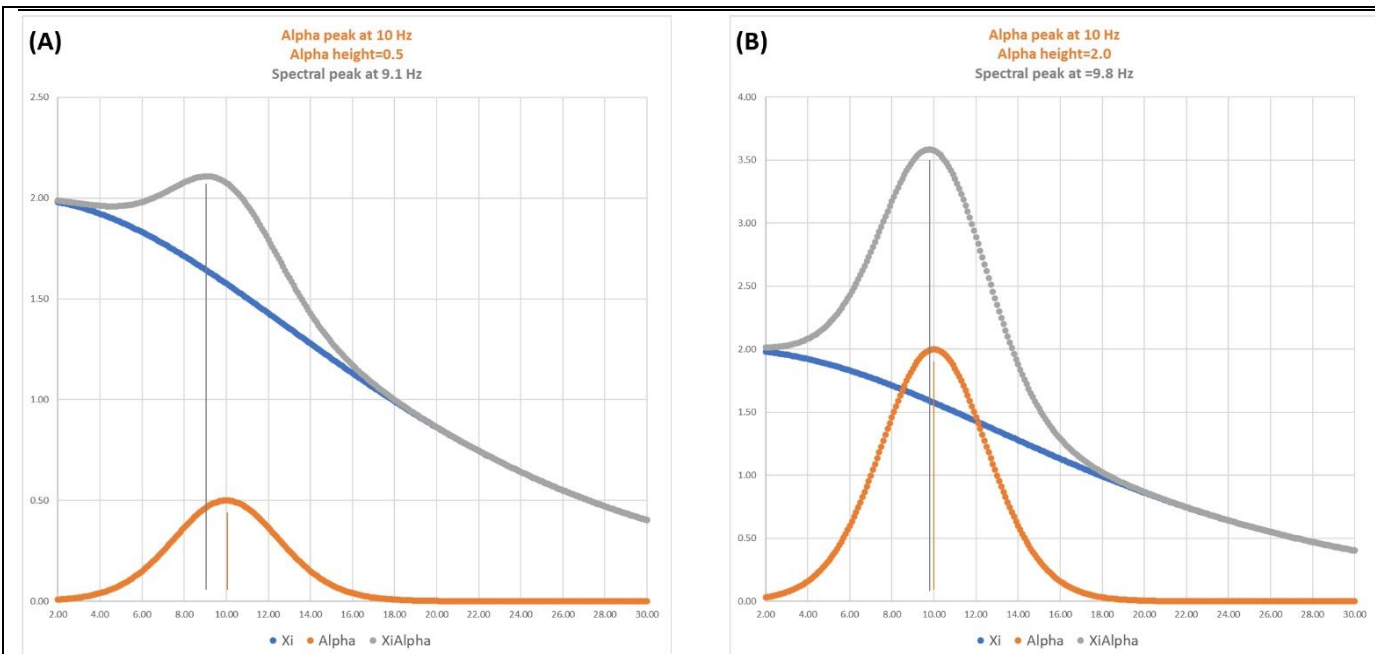

Figure 2: Spectral density $\phi(\omega)$ as a function of discrete frequency (Hz). Eq. 2 Xi: blue; Eq. 3 Alpha: orange; Eq. 1 Xi-Alpha: grey. Parameters common to both (A) and (B): b=2; c=0.001; d=2.5; f=0.008; g=10; $\omega_\alpha$=10. Alpha amplitude parameter with value e=0.5 for (A); e=2 for (B). Despite same Alpha generator peak frequency at 10 Hz, the individual alpha peak frequencies (IAF) for the Xi-Alpha spectra are different: IAF=9.1 Hz for (A); IAF=9.9 Hz for (B).

The result shown in Figure 2 illustrates for a fact that even with one single Xi process and one single alpha process common to the whole cortex, the Xi-Alpha model can explain a rich repertoire of diverse spatio-spectral properties by varying only the amplitudes.

### 11.B. Empirical separation of the cortical Xi and the Alpha processes by non-negative matrix factorization (NMF)

Empirical evidence supporting the whole cortex Xi-Alpha model is shown in Figure 3. First, the average scaled (see Appendix A) EEG cross-spectra of 203 participants in eyes closed condition (Babayan et al (2019)) was computed, followed by estimating the eLORETA cortical spectral power values at 6239 grey matter voxels (diagonal elements from Eq. 33), at 44 discrete frequencies (frequency range 2-44 Hz).

The non-negative matrix factorization of this data resulted in 99.62% of explained variance for two components, which are shown in Figure 3.

Further empirical evidence supporting the whole cortex Xi-Alpha model is shown in Figure 4. Using a completely independent EEG database, in this case from van Dijk et al (2022), consisting of 47 participants in eyes closed resting state condition, the same non-negative matrix factorization procedure as previously described was applied here, which resulted in 99.15% of explained variance for two components, which are shown in Figure 4.

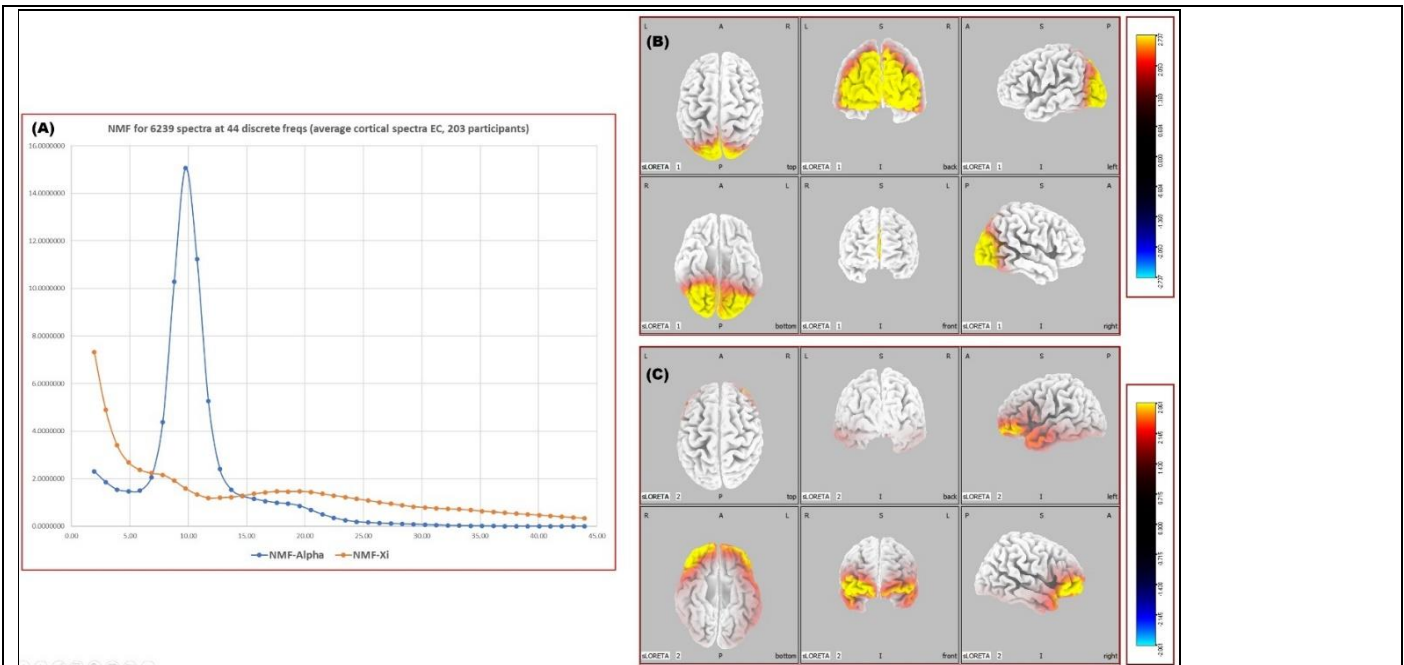

Figure 3: Non-negative matrix factorization of eLORETA cortical spectral power values at 6239 grey matter voxels, at 44 discrete frequencies (frequency range 2-44 Hz), obtained from the average EEG cross-spectra of 203 participants (Babayan et al 2019) in eyes closed condition. Two components produced 99.62% explained variance. (A): the two spectral components which are unequivocally identified as the Xi and the Alpha processes. (B): the Alpha cortical activity distribution; (C): the Xi cortical activity distribution.

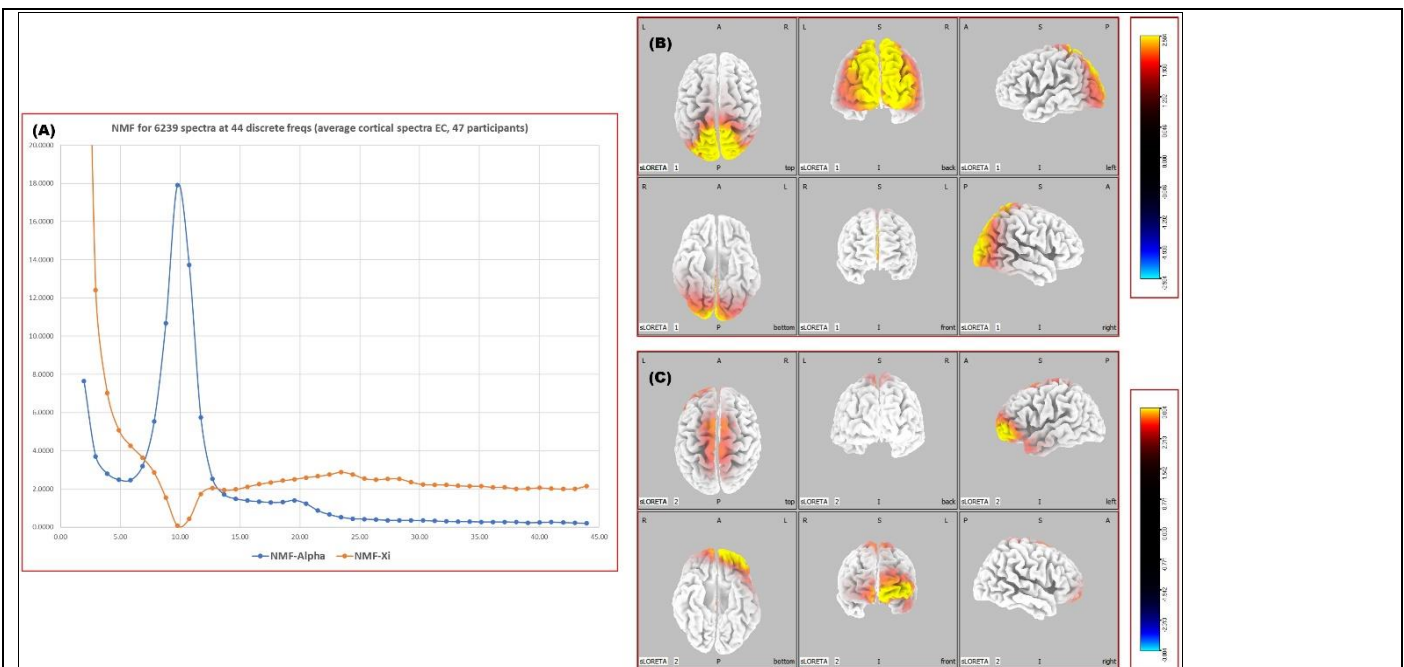

Figure 4: Non-negative matrix factorization of eLORETA cortical spectral power values at 6239 grey matter voxels, at 44 discrete frequencies (frequency range 2-44 Hz), obtained from the average EEG cross-spectra of 47 participants (van Dijk et al 2022) in eyes closed condition. Two components produced 99.15% explained variance. (A): the two spectral components which are unequivocally identified as the Xi and the Alpha processes. (B): the Alpha cortical activity distribution; (C): the Xi cortical activity distribution.

The non-negative matrix factorization analyses of different resting state EEG databases are in remarkably good qualitative agreement, even though EEG hardware, preprocessing pipelines, number of electrodes, and number of participants are quite different. Both results indicate that two non-negative components suffice, and that the components can be unequivocally assigned to the Xi process with decreasing power and more strongly located towards frontal cortices, and to the Alpha process with a spectral peak at around 10 Hz and more strongly located towards posterior cortices.

These results lend strong and compelling support to the existence of the two processes, Xi and Alpha, in human resting state cortical activity, especially because the method is purely data driven, and does not force any a given form for the spectra.

### 11.C. The estimated cortical Xi-Alpha model: spectral aspects

Figure 5 shows the actual data, corresponding to the global field power spectra, i.e. the trace of the scaled (see Appendix A), of the average reference EEG cross-spectral matrices. The frequency-by-frequency medians of spectral power values for the two data sets (N=203, N=47) in two conditions (EC, EO) are shown in Figure 5A, while Figure 5B, 5C, 5D, 5E show the median for each data set and each condition separately, accompanied by the 25 and 75 percentiles, which provide robust information on the variances of the spectra.

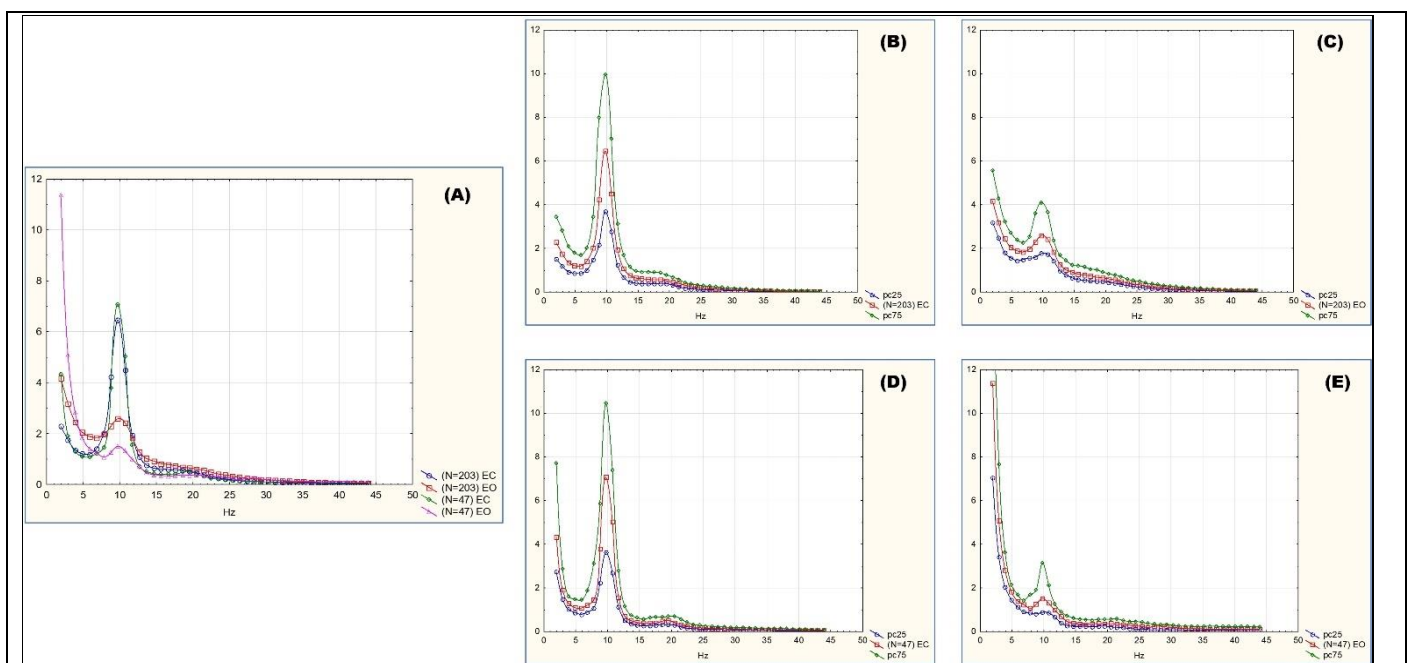

Figure 5: Actual data, corresponding to the global field power spectra, i.e. the trace of the EEG cross-spectral matrices. The frequency-by-frequency medians of spectral power values for the two data sets (N=203, N=47) in two conditions (EC, EO) are shown in Figure 5A, while Figure 5B, 5C, 5D, 5E show the median for each data set and each condition separately, accompanied by the 25 and 75 percentiles, which provide robust information on the variances of the spectra.

Figure 6 shows the best fit Xi-Alpha model (Eq. 18) to the global field power spectral data, i.e. to the trace of the EEG cross-spectral matrices (based on (203x2)+(47x2) distinct least squares minimizations). The frequency-by-frequency medians of spectral power values for the best fits to two data sets (N=203, N=47) in two conditions (EC, EO) are shown in Figure 6A, while Figure 6B, 6C, 6D, 6E show the median for the best fit to each data set and each condition separately, accompanied by the 25 and 75 percentiles, which provide robust information on the variances of the fitted spectra.

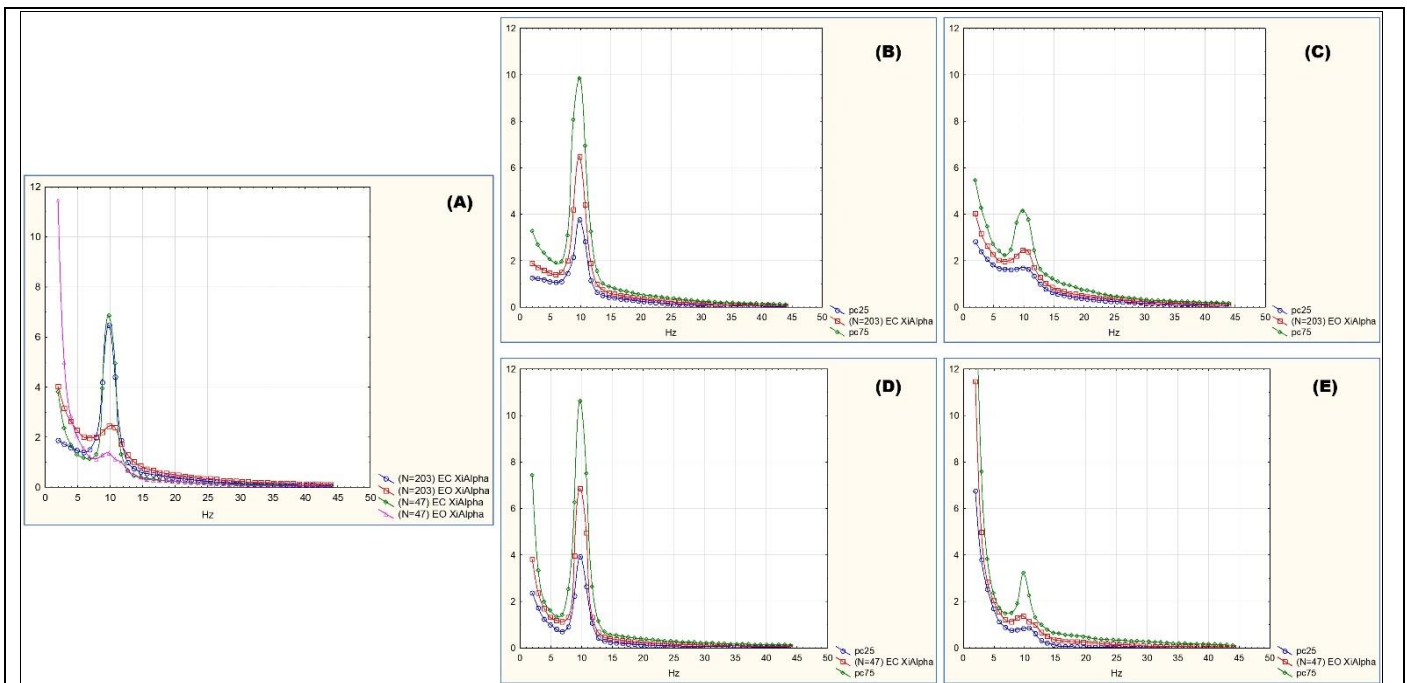

Figure 6: Best fit Xi-Alpha model (Eq. 18) to the global field power spectral data, i.e. to the trace of the EEG cross-spectral matrices (based on (203x2)+(47x2) least squares minimizations). The frequency-by-frequency median spectral power values for the best fits to two data sets (N=203, N=47) in two conditions (EC, EO) are shown in Figure 6A, while Figure 6B, 6C, 6D, 6E show the median for the best fit to each data set and each condition separately, accompanied by the 25 and 75 percentiles, which provide robust information on the variances of the fitted spectra.

Figure 7 shows the median data and the median best fit, for global field power spectra (see Figure 5A and Figure 6A).

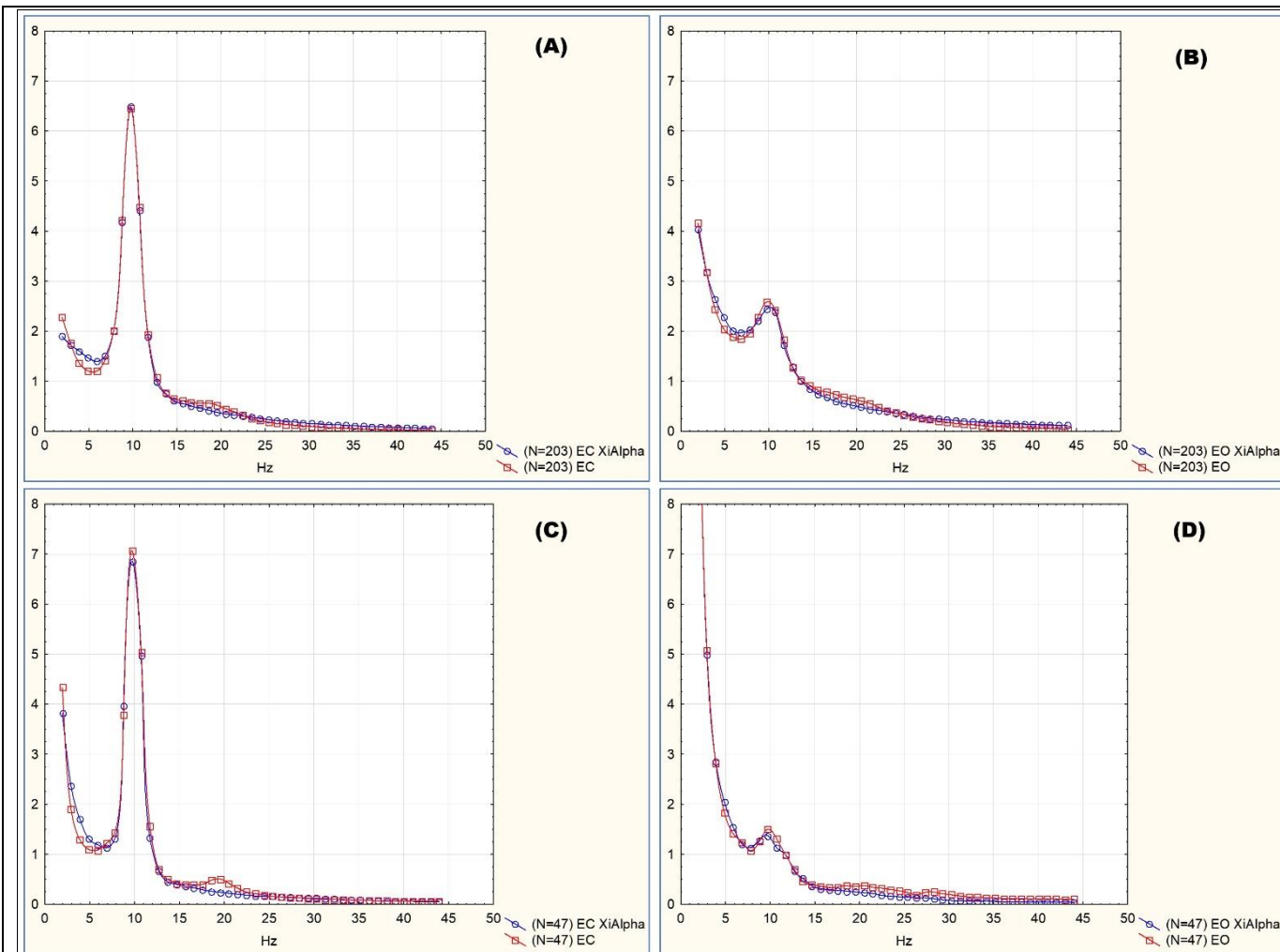

Figure 7: The median data and the median best fit, for global field power spectra (see Figure 6A and Figure 5A), for two data sets (N=203, N=47) in two conditions (EC, EO).

The median values of the parameters estimated for the Xi-Alpha model (Eq. 18) for two EEG data sets under two different conditions are listed in Table 1.

| Table 1: Parameters and goodness of fit for the Xi-Alpha model | | | | | | | | | | | | |
|---|---|---|---|---|---|---|---|---|---|---|---|---|
| | N=203, EC | | | N=203, EO | | | N=47, EC | | | N=47, EO | | |
| | 35 %tile | Median | 65 %tile | 35 %tile | Median | 65 %tile | 35 %tile | Median | 65 %tile | 35 %tile | Median | 65 %tile |
| **%ExpVar FullMatrix** | 94.41800 | 95.95500 | 97.05700 | 92.94100 | 94.66000 | 95.60800 | 94.02600 | 95.54900 | 96.49400 | 94.43400 | 96.37000 | 98.26300 |
| **Dim** | 1.71300 | 1.77700 | 1.85800 | 1.63800 | 1.72100 | 1.81600 | 2.19600 | 2.26300 | 2.43700 | 2.44800 | 2.58800 | 2.80500 |
| **XiHeight** | 1.53685 | 2.02082 | 3.11830 | 4.29814 | 5.85127 | 7.42535 | 9.89818 | 21.60817 | 70.88062 | 34.26173 | 65.15668 | 87.99560 |
| **XiNarrow** | 0.00118 | 0.00516 | 0.03367 | 0.03153 | 0.08109 | 0.13748 | 0.65324 | 2.07843 | 3.77041 | 0.25099 | 0.46559 | 3.15960 |
| **XiExponent** | 0.91926 | 1.72615 | 3.76013 | 0.68726 | 0.77866 | 1.02589 | 0.59708 | 0.72791 | 1.01799 | 0.78322 | 1.21897 | 1.55084 |
| **AlphaHeight** | 5.31922 | 7.56977 | 9.70516 | 1.42244 | 2.04023 | 2.90055 | 6.39888 | 8.43800 | 9.82801 | 0.98008 | 1.70876 | 3.75347 |
| **AlphaNarrow** | 0.02340 | 0.02842 | 0.03341 | 0.00710 | 0.01641 | 0.02112 | 0.02282 | 0.02947 | 0.03884 | 0.01373 | 0.02183 | 0.02820 |
| **AlphaHz** | 9.52034 | 9.90362 | 10.25355 | 9.77623 | 10.17033 | 10.75537 | 9.68751 | 9.87207 | 10.08255 | 9.80257 | 10.23716 | 10.27624 |
| **%ExpVar Spectra** | 98.55847 | 99.27972 | 99.57418 | 98.61466 | 99.18821 | 99.41247 | 98.20102 | 98.70001 | 99.07872 | 98.75321 | 99.29058 | 99.53568 |
| EC: eyes closed. EO: eyes open. N=203 EEGs from Babayan et al (2019). N=47 EEGs from van Dijk et al (2022). Median values with percentile values at 35% and 65% are listed. Dim: number of components from Eq. 37. From Eq. 18: XiHeight=b, XiNarrow=c, XiExponent=d, AlphaHeight=e, AlphaNarrow=f, AlphaHz=$\omega_\alpha$, fixed alpha exponent g=20. [%ExpVar FullMatrix] given by Eq. 25. [%ExpVar Spectra] given by Eq. 22. | | | | | | | | | | | | |

The median of the goodness of fit for the full cross-spectral Xi-Alpha model is in the range 94.7% to 96.4% of explained variance (Eq. 25). The XI-Alpha model for the trace of the cross-spectral matrices (i.e. global field power spectra) has a median goodness of fit (Eq. 22) in the range 98.7% to 99.3% of explained variance.

The estimated number of cortical processes based on entropy (Eq. 37) lies in the range 1.72 to 2.59. This justifies two processes, namely the Xi and the Alpha processes, as adequate for describing resting state cortical electric neuronal activity.

Admittedly, there are obvious differences between the EEG cross-spectra of the two data sets (Figure 5), which is noted from Table 1 with respect to the Xi-Alpha parameters. But in all cases, the Xi-Alpha model fits extremely well each data set and condition, as shown in Table 1, and seen from comparing Figure 5 with Figure 6, and as shown in Figure 7.

It is not the aim of the study to uncover the reasons for the differences between the data sets, which may be partly caused by differences in recording hardware and their settings, preprocessing pipelines of the EEGs, number of subjects, and instructions given to subjects prior and during the recordings.

Despite the quantitative detailed discrepancies between the two data sets, they both have in common that they are very well described by the two main processes: Xi and Alpha.

The multichannel EEG Xi-Alpha model is capable of very high goodness-of-fit, including not only the spectral shape, but also the full Hermitian covariance structure at all frequencies in the range studied here, namely 2 Hz to 44 Hz.

There are at least two distinct approaches in making use of the spectral properties of the Xi and Alpha processes, when comparing one participant against a normative group, or when comparing two groups of participants (e.g. healthy controls and some disorder), or when comparing the same group under two conditions (e.g. rest against mental arithmetic), or when tasked with discovering subgroups (e.g. cluster analysis):
1. Use the set of Xi-Alpha spectral parameters (those in Eq. 2 and Eq. 3, in Table 1).
2. Use the set of values for $\left[\xi(\omega), \alpha(\omega)\right]$ at all discrete frequencies $\omega = 1...N_\Omega$ evaluated from Eq. 2 and Eq. 3 with the best fitting parameters (Figure 8, see also Figure 6 and Figure 7).

Figure 8 illustrates the best fit Xi process and best fit Alpha process (the two additive terms in the right hand side of Eq. 18), for the Babayan et al (2019) eyes closed EEG data set. Medians and percentiles at 25% and 75% are shown, calculated at each discrete frequency ($\omega = 1...N_\Omega$), for 203 participants. Such type of data can be used for statistical analyses, as mentioned in the previous paragraph.

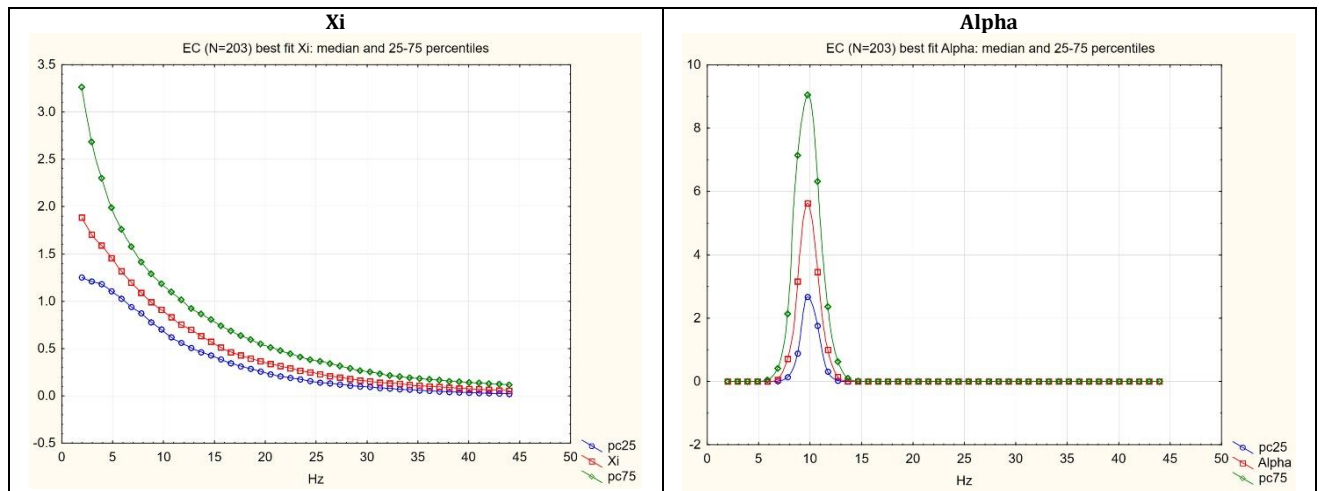

Figure 8: Best fit Xi process and best fit Alpha process (the two additive terms in the right hand side of Eq. 18), for the Babayan et al (2019) eyes closed EEG data set. Medians and percentiles at 25% and 75% are shown, calculated at each discrete frequency, for 203 participants.

### 11.D. Functional images of the cortical Xi and Alpha processes

Figure 9 (for dataset N=203, Babayan et al 2019) and Figure 10 (for data set N=47, van Dijk et al 2022) display the cortical generator distribution, given by the voxel-by-voxel median of spectral power values across all participants, for the Xi and the Alpha processes, based on eLORETA computations in Eq. 27. Note that the diagonal elements of the matrices in the left-hand side of Eq. 27 provide the cortical generator distribution for each process, which was computed for each dataset and each condition and each subject.

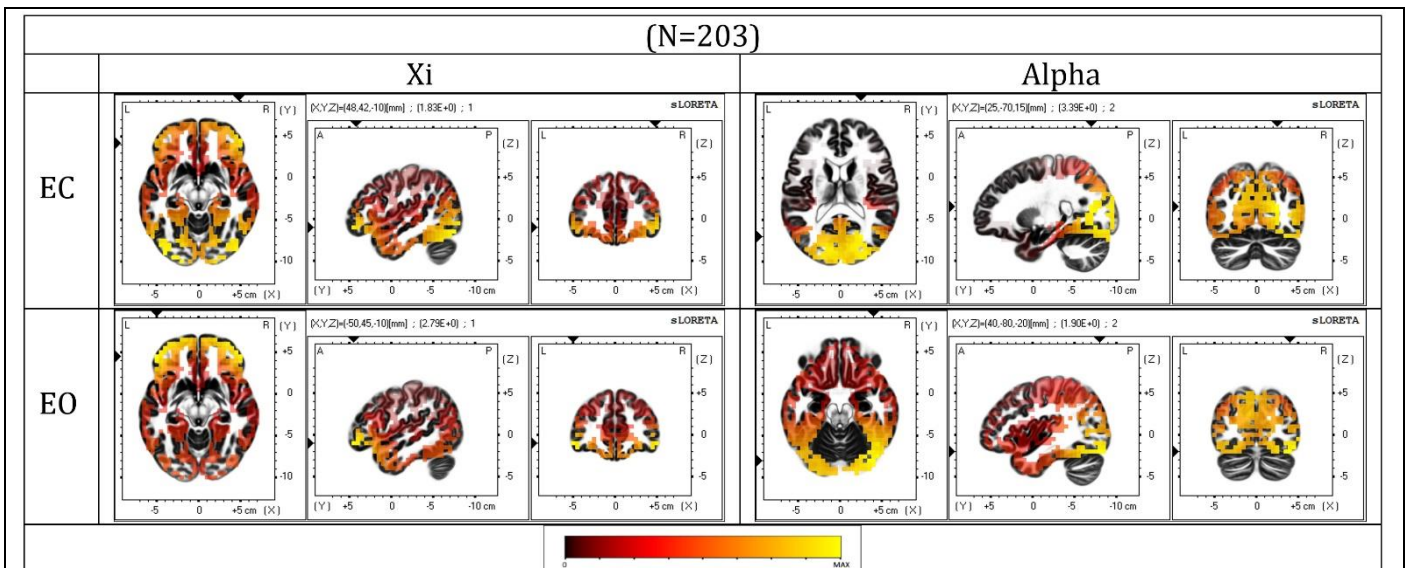

Figure 9: Cortical generator distributions for dataset Babayan et al 2019, given by the voxel-by-voxel median of spectral power values across all participants (N=203), for eyes closed (EC) and eyes open (EO) conditions, for the Xi and the Alpha processes, based on eLORETA computations in Eq. 27. Note that the diagonal elements of the matrices in the left-hand side of Eq. 27 provide the cortical generator distribution for each process, which was computed for each condition and each subject.

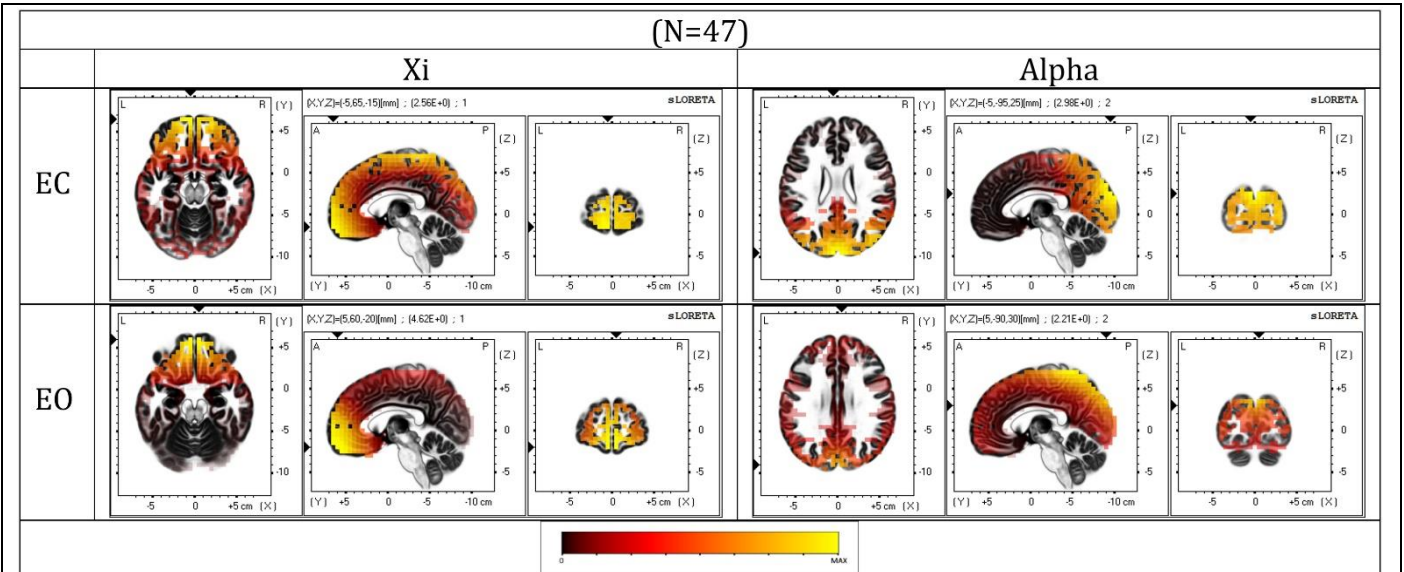

Figure 10: Cortical generator distributions for dataset van Dijk et al 2022, given by the voxel-by-voxel median of spectral power values across all participants (N=47), for eyes closed (EC) and eyes open (EO) conditions, for the Xi and the Alpha processes, based on eLORETA computations in Eq. 27. Note that the diagonal elements of the matrices in the left-hand side of Eq. 27 provide the cortical generator distribution for each process, which was computed for each condition and each subject.

The notable qualitative features are:
1. The Xi process has strongest generators located in frontal cortical areas, which are strongest in EO condition. This holds for the two independent datasets.
2. The Alpha process has strongest generators located in posterior cortical areas, which are strongest in EC condition. This holds for the two independent datasets.

### 11.E. The estimated cortical connectivity properties of the Xi-Alpha model

An exploratory analysis of the cortical connectivity properties of the Xi-Alpha model follows. First, the average scaled (see Appendix A) EEG cross-spectra of 203 participants in eyes closed condition (Babayan et al (2019)) was computed and is denoted as $\left[\hat{\mathbf{\Phi}}(\omega)\right]$. This population EEG cross-spectrum was used (as explained in sections "6: Estimation outline of the cortical Xi-Alpha model" and section "6.A: Cortical activity and intracortical connectivity") for the estimation of $\left[\hat{\mathbf{R}}_\xi, \hat{\mathbf{R}}_\alpha\right]$, which denote the Hermitian coherence matrices for each process. Note that the coherences of the Xi process and of the Alpha process $\left[\hat{\mathbf{R}}_\xi, \hat{\mathbf{R}}_\alpha\right]$ do not change with frequency. These are very large matrices, in this case 6239x6239. From here, the coherences for only a subset of 61 cortical regions of interest (ROIs) were used for further analysis, corresponding to the 61 cortical voxels closest to each scalp electrode used in the Babayan et al (2019) EEG recordings. The MNI-coordinates and detailed neuroanatomical information of the 61 ROIs are given in Appendix B.

Note that the analysis here is based on cortical coherences, i.e. it is emphatically and unequivocally NOT based on coherences of scalp EEG signals. The only relation to scalp EEG electrodes is in the use of "cortical voxels underlying scalp EEG electrodes".

Figure 11 shows scatter plots of log lagged coherences as a function of interdistance, for 1830 values, i.e. all distinct pairs among 61 cortical ROIs:

Eq. 41
$$\left\{ \begin{array}{l} \ln\left( \dfrac{\left| \text{Im}\left[ \hat{\mathbf{R}}_\xi \right]_{uv} \right|}{\sqrt{1-\left(\text{Re}\left[ \hat{\mathbf{R}}_\xi \right]_{uv}\right)^2}} \right) \textit{ versus } (d_{uv}) \\ \ln\left( \dfrac{\left| \text{Im}\left[ \hat{\mathbf{R}}_\alpha \right]_{uv} \right|}{\sqrt{1-\left(\text{Re}\left[ \hat{\mathbf{R}}_\alpha \right]_{uv}\right)^2}} \right) \textit{ versus } (d_{uv}) \end{array} \right\}$$

where $\hat{\mathbf{R}}_\xi$ and $\hat{\mathbf{R}}_\alpha$ denote the estimated coherence matrices for the Xi and the Alpha process (see section 6.A: Cortical activity and intracortical connectivity), $\left[ \hat{\mathbf{R}} \right]_{uv}$ denotes its complex valued element for the $u$-th and $v$-th voxel pair, Re and Im denote the real and imaginary parts of a complex number, and $d_{uv}$ is the Euclidean distance between the $u$-th and $v$-th cortical voxel pair.

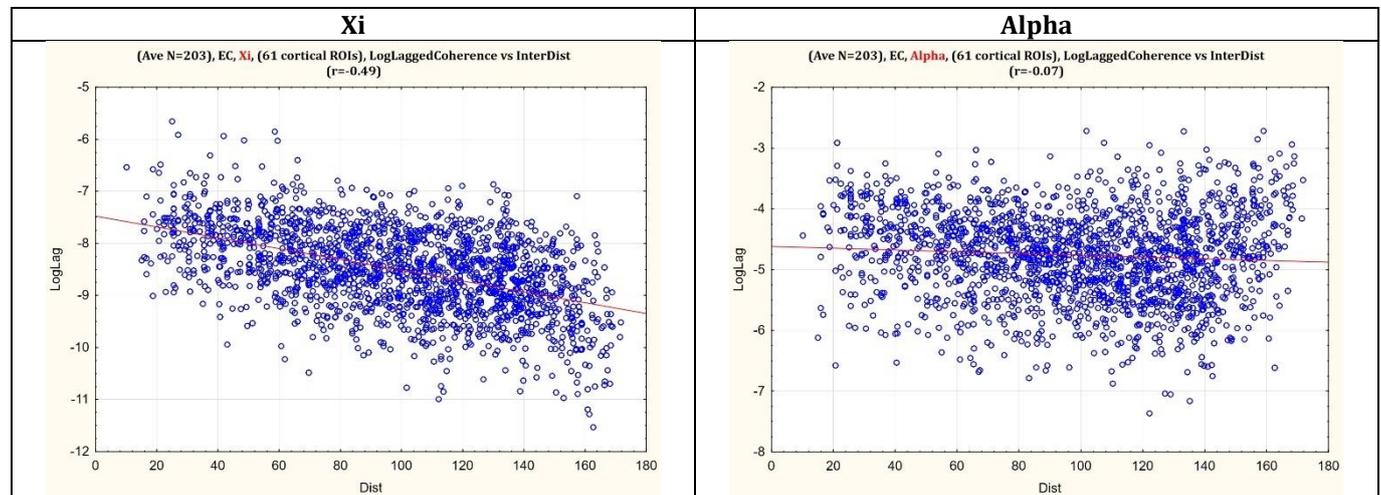

Figure 11: Scatter plots of log lagged coherences of the Xi and the Alpha processes, as a function of interdistance, for 1830 values, i.e. all distinct pairs among 61 cortical ROIs. See text for details.

Note that the Xi process has lagged coherences that decrease with interdistance very significantly (correlation coefficient=-0.49 for sample size 1830), with very near large effect size. However, the Alpha process has very low correlation coefficient=-0.07, which does not even have a small effect size.

These results characterize the Xi process as having intracortical finite time information flow in the form of an isotropic spatial process, for which the lagged connectivity strength between two points is a monotonically decreasing function of interdistance. No such structure was observed for the Alpha process.

In Pascual-Marqui et al (1988) the Xi process for scalp EEG was shown to have isotropic characteristics. The new results presented here support a similar property for the actual Xi cortical generators.

### 11.F. The laminar distribution of the Xi and Alpha processes

Consider the analysis of cortical spectra using non-negative matrix factorization, detailed in section "7: Exploratory method for the discovery of processes based on non-negative matrix factorization (NMF)". Refer to the spectra for the two non-negative factors that were empirically identified as the Xi and the Alpha processes (Figures 3A and 4A). The relative Xi and the relative Alpha cortical spectra are defined as:

**Eq. 42** $$\xi_{Rel}(\omega) = \frac{\xi(\omega)}{\xi(\omega) + \alpha(\omega)}$$

**Eq. 43** $$\alpha_{Rel}(\omega) = \frac{\alpha(\omega)}{\xi(\omega) + \alpha(\omega)}$$

Figure 12 displays the relative spectra corresponding to the NMF-based cortical spectra of Figures 3A and 4A.

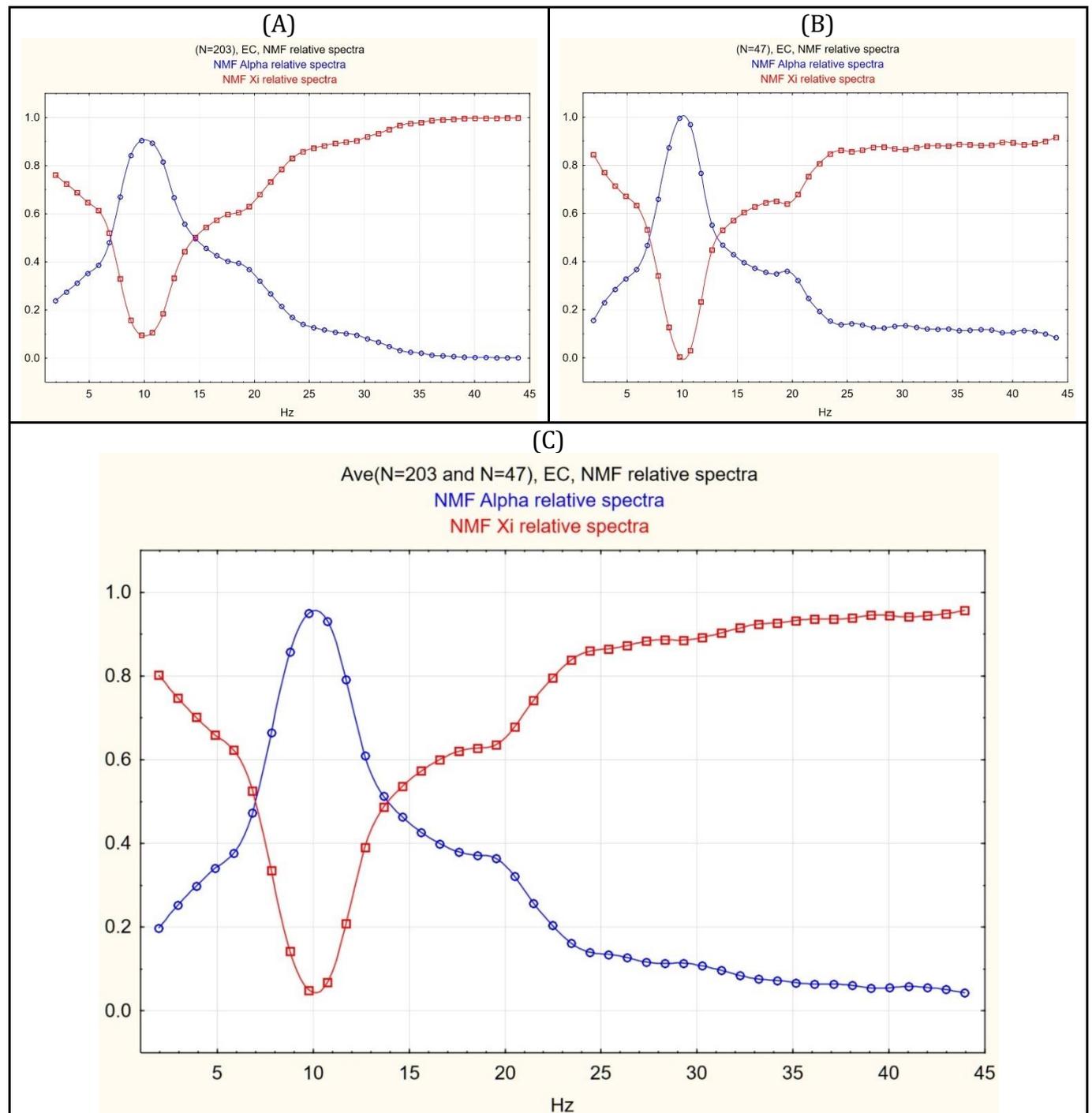

Figure 12: Relative spectra corresponding to the NMF-based cortical spectra of Figures 3A and 4A, estimated from Eq. 42 and Eq. 43. (A): Relative NMF cortical spectra for the eyes closed Babayan et al (2019) dataset of Figure 3A. (B): Relative NMF cortical spectra for the eyes closed Dijk et al (2022) dataset of Figure 4A. (C): Average of relative cortical spectra in (A) and (B).

There is a striking qualitative similarity between Figure 12 here, and Figures 1d and 1g from Mendoza-Halliday et al (2022), which are the basis for their main result: "…common spectrolaminar pattern characterized by an increasing deep-to-superficial layer gradient of gamma frequency LFP power peaking in layers 2/3, and an increasing superficial-to-deep gradient of alpha-beta power peaking in layers 5/6."

In Mendoza-Halliday et al (2022) the relative powers were computed from intracortical electrophysiological recordings using multicontact laminar probes (16, 24, or 32 contacts). This can be taken as ground truth.

On the other hand, applying an inverse imaging method such as eLORETA to non-invasive EEG recordings will at best provide a low-resolution estimator for the mixture of spectra across all cortical layers. By applying a "blind" unmixing procedure, such as NMF, two components are obtained which can be empirically assigned as follows (based on Mendoza-Halliday et al (2022)):
1. The Alpha process corresponding to generators in layers 5/6.
2. The Xi process corresponding to generators in layers 2/3.

Model based (Eq. 11) relative cortical spectra for the Xi and Alpha processes are shown in Figure 13, corresponding to frequency-by-frequency medians over 203 best fitting Xi and Alpha spectra, for eyes closed data from Babayan et al (2019), together with percentiles at 25% and 75%. Very similar results, not shown here, are obtained from the van Dijk et al (2022) dataset.

Note the qualitative agreement between the model-based relative spectra in Figure 13 and the empirical NMF-based relative spectra in Figure 12.

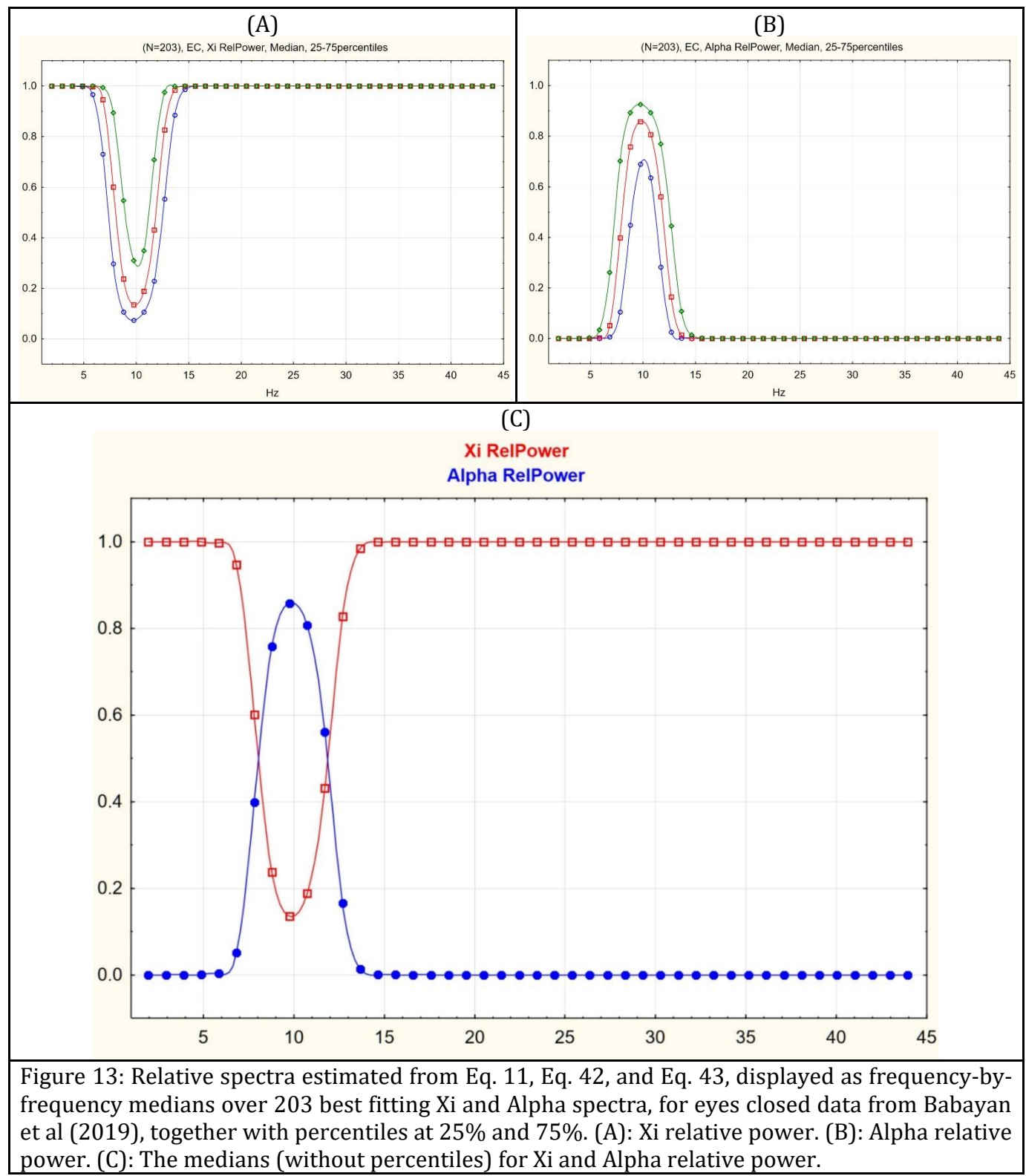

Figure 13: Relative spectra estimated from Eq. 11, Eq. 42, and Eq. 43, displayed as frequency-by-frequency medians over 203 best fitting Xi and Alpha spectra, for eyes closed data from Babayan et al (2019), together with percentiles at 25% and 75%. (A): Xi relative power. (B): Alpha relative power. (C): The medians (without percentiles) for Xi and Alpha relative power.

## 12. Concluding summary

Two independent open-access, resting state eyes open and closed EEG data sets (203 participants with 61 electrodes, and 47 participants with 26 electrodes) were used to demonstrate, test, and validate the cortical Xi-Alpha model.

The median value of explained variance was 95% for the "cortical Xi-Alpha model" for the full cross-spectra, across all subjects in all datasets and conditions.

The average dimension of the vectorized cross-spectra lies between two and three for the whole cortex, justifying two processes as an adequate approximation for the resting state EEG. However, more than two processes can be accommodated by the model.

A non-negative matrix factorization analysis of population power spectra sampled at 6239 cortical grey matter voxels with only two components explains 99% of the variance. The spectral shapes and the 3D cortical activity images unequivocally are identified as the Xi and Alpha processes.

The Alpha process is more strongly located in posterior cortical regions, while the Xi process is more distributed and leans towards frontal regions.

The Xi lagged connectivity matrix for cortical sources is isotropic with interdistance, while Alpha is not.

Recent results from laminar recordings by Mendoza-Halliday et al (2022) justify the attribution of pyramidal neurons in layers 2/3 as generators of the Xi-process, and pyramidal neurons in layers 5/6 as generators of the Alpha process.

## 13. Appendix A: Estimation of the "cortical Xi-Alpha model"

Average reference EEG is used throughout.

EEG cross-spectral matrices are estimated in this study as the average periodogram over EEG epochs, computed via the discrete Fourier transform using a Hann taper (see e.g. Brillinger (2001), Oppenheim and Schafer (2014), Frei et al (2001), Pascual-Marqui et al (2021)).

Let $\hat{\Phi}(\omega) \in \mathbb{C}^{N_E \times N_E}$ denote the EEG cross-spectral matrix for $N_E$ electrodes, at discrete frequency $\omega$, with $\omega = 1 \ldots N_\Omega$, with $N_\Omega$ denoting the number of discrete frequencies. Let:

Eq. 44 $\quad GFP = \dfrac{1}{N_\Omega N_E} \sum_{i=1}^{N_E} \sum_{\omega=1}^{N_\Omega} \left[ \hat{\Phi}(\omega) \right]_{ii}$

be the average Global Field Power (GFP) over electrodes and frequencies.

Unless otherwise stated, scaled cross-spectra are used throughout this study in the analyses:

Eq. 45 $\quad \hat{\Phi}(\omega) \leftarrow \dfrac{1}{GFP} \hat{\Phi}(\omega)$

Subject to subject EEG cross-spectra is characterized by very high variability in GFP, which is accounted for and eliminated with the scaled cross-spectra (Eq. 45).

Consider the least squares functional in Eq. 20 with $\hat{u}(\omega)$ given by Eq. 21 and Eq. 45, and with $u(\omega)$ given by Eq. 18, with parameters denoted as $p_i$, with $i=1...6$:

Eq. 46
$$\begin{cases} p_1 = (\tau_\xi b) \ ; \ p_2 = c \ ; \ p_3 = d \ ; \ p_4 = (\tau_\alpha e) \ ; \ p_5 = f \ ; \ p_6 = \omega_\alpha \\ fixed \ g = 20 \end{cases}$$

Newton's method for optimization (see e.g. Givens and Hoeting (2013), Equation 2.9 therein) was used, one parameter at a time, as shown with pseudo-code in Box 1, with initial values:

Eq. 47
$$\begin{cases} p_1 = (\tau_\xi b) = 9 \ ; \ p_2 = c = 0.04 \ ; \ p_3 = d = 1 \ ; \ p_4 = (\tau_\alpha e) = 6 \ ; \ p_5 = f = 0.1 \ ; \ p_6 = \omega_\alpha = 10 \\ fixed \ g = 20 \end{cases}$$

```
repeat1
  for i:=1 to 6 do begin
    repeat2
```
$$p_i \leftarrow \left[ p_i - \left| \frac{\partial^2 F}{\partial p_i^2} \right|^{-1} \left( \frac{\partial F}{\partial p_i} \right) \right]$$
```
    until convergence2
  end
until convergence1
```

Box 1: Pseudo-code for estimating the Xi-Alpha parameters. *F*: functional (Eq. 20). See Eq. 18, Eq. 21, Eq. 45, Eq. 46. Convergence in both "repeats" occurs when the number of iterations is greater than or equal to 10'000, or when $|1-(F_{old}/F_{new})| < 10^{-5}$, where $F_{old}$ and $F_{new}$ are the values of the functional in the previous and current iteration. This is applied continuously withing both repeat cycles.

At this point, estimators for the Xi-Alpha spectral parameters and the actual Xi and Alpha spectra (Eq. 2 and Eq. 3), which are common to both scalp EEG and cortical electric neuronal activity are available.

Given estimators for the Xi-Alpha parameters (Box 1), the solution to the least squares functional in Eq. 24 for the Xi and Alpha Hermitian covariances is:

Eq. 48
$$\begin{cases} \hat{\mathbf{\Phi}}_\xi = \dfrac{\overline{(a^2)}\overline{(x\mathbf{D})} - \overline{(xa)}\overline{(a\mathbf{D})}}{\overline{(a^2)}\overline{(x^2)} - \overline{(xa)}^2} \\ \\ \hat{\mathbf{\Phi}}_\alpha = \dfrac{\overline{(x^2)}\overline{(a\mathbf{D})} - \overline{(xa)}\overline{(x\mathbf{D})}}{\overline{(x^2)}\overline{(a^2)} - \overline{(xa)}^2} \end{cases}$$

with:

Eq. 49
$$\begin{cases} \overline{(a^2)} = \dfrac{1}{N_\Omega} \sum_{\omega=1}^{N_\Omega} \hat{\alpha}^2(\omega) \ ; \ \overline{(x^2)} = \dfrac{1}{N_\Omega} \sum_{\omega=1}^{N_\Omega} \hat{\xi}^2(\omega) \ ; \ \overline{(xa)} = \dfrac{1}{N_\Omega} \sum_{\omega=1}^{N_\Omega} \hat{\xi}(\omega)\hat{\alpha}(\omega) \\ \overline{(x\mathbf{D})} = \dfrac{1}{N_\Omega} \sum_{\omega=1}^{N_\Omega} \hat{\xi}(\omega)\hat{\mathbf{\Phi}}(\omega) \ ; \ \overline{(a\mathbf{D})} = \dfrac{1}{N_\Omega} \sum_{\omega=1}^{N_\Omega} \hat{\alpha}\hat{\mathbf{\Phi}}(\omega) \end{cases}$$

If only Xi is present (absence of the Alpha process), then:

Eq. 50 $\quad \hat{\mathbf{\Phi}}_\xi = \dfrac{\overline{(x\mathbf{D})}}{\overline{(x^2)}}$

Note: since Eq. 48 does not guarantee positive definiteness for $\hat{\mathbf{\Phi}}_\xi$ and $\hat{\mathbf{\Phi}}_\alpha$, a final step consists of using their singular value decompositions, and reconstruction with only positive eigenvalues. These final estimators are used for declaring the global least squares fit value in Eq. 24.

Finally, the Xi and Alpha Hermitian covariance matrices for cortical electric neuronal activity $\left[\hat{\mathbf{\Psi}}_\xi, \hat{\mathbf{\Psi}}_\alpha\right]$ are estimated using a linear inverse solution, such as eLORETA, by means of Eq. 27.

# 14. Appendix B: MNI coordinates for 61 ROIs used in connectivity analysis of the cortical Xi-Alpha model

| ROI# | X-MNI | Y-MNI | Z-MNI | Lobe | Structure | Brodmann area |
| --- | --- | --- | --- | --- | --- | --- |
| 1 | -25 | 65 | -5 | Frontal Lobe | Superior Frontal Gyrus | BA10 |
| 2 | 25 | 65 | -5 | Frontal Lobe | Superior Frontal Gyrus | BA10 |
| 3 | -50 | 40 | -10 | Frontal Lobe | Inferior Frontal Gyrus | BA47 |
| 4 | -40 | 45 | 30 | Frontal Lobe | Middle Frontal Gyrus | BA10 |
| 5 | 5 | 45 | 50 | Frontal Lobe | Superior Frontal Gyrus | BA8 |
| 6 | 40 | 45 | 30 | Frontal Lobe | Middle Frontal Gyrus | BA10 |
| 7 | 50 | 40 | -10 | Frontal Lobe | Inferior Frontal Gyrus | BA47 |
| 8 | -60 | 15 | 20 | Frontal Lobe | Inferior Frontal Gyrus | BA45 |
| 9 | -20 | 15 | 65 | Frontal Lobe | Superior Frontal Gyrus | BA6 |
| 10 | 30 | 10 | 65 | Frontal Lobe | Middle Frontal Gyrus | BA6 |
| 11 | 60 | 15 | 20 | Frontal Lobe | Inferior Frontal Gyrus | BA44 |
| 12 | -65 | -15 | -15 | Temporal Lobe | Middle Temporal Gyrus | BA21 |
| 13 | -50 | -20 | 60 | Parietal Lobe | Postcentral Gyrus | BA3 |
| 14 | 5 | -10 | 70 | Frontal Lobe | Superior Frontal Gyrus | BA6 |
| 15 | 55 | -20 | 55 | Parietal Lobe | Postcentral Gyrus | BA1 |
| 16 | 70 | -20 | -10 | Temporal Lobe | Middle Temporal Gyrus | BA21 |
| 17 | -65 | -45 | 30 | Parietal Lobe | Supramarginal Gyrus | BA40 |
| 18 | -30 | -45 | 70 | Parietal Lobe | Postcentral Gyrus | BA5 |
| 19 | 30 | -45 | 70 | Parietal Lobe | Postcentral Gyrus | BA5 |
| 20 | 65 | -50 | 30 | Parietal Lobe | Supramarginal Gyrus | BA40 |
| 21 | 5 | 60 | 30 | Frontal Lobe | Superior Frontal Gyrus | BA10 |
| 22 | -60 | -65 | -10 | Temporal Lobe | Inferior Temporal Gyrus | BA37 |
| 23 | -40 | -70 | 45 | Parietal Lobe | Inferior Parietal Lobule | BA7 |
| 24 | -5 | -65 | 65 | Parietal Lobe | Precuneus | BA7 |
| 25 | 45 | -70 | 45 | Parietal Lobe | Inferior Parietal Lobule | BA7 |
| 26 | 55 | -70 | 0 | Occipital Lobe | Middle Occipital Gyrus | BA37 |
| 27 | -35 | -90 | -20 | Occipital Lobe | Inferior Occipital Gyrus | BA18 |
| 28 | -20 | -100 | 10 | Occipital Lobe | Middle Occipital Gyrus | BA19 |
| 29 | -5 | -100 | 15 | Occipital Lobe | Cuneus | BA18 |
| 30 | 20 | -100 | 5 | Occipital Lobe | Middle Occipital Gyrus | BA18 |
| 31 | 40 | -85 | -20 | Occipital Lobe | Inferior Occipital Gyrus | BA18 |
| 32 | -40 | 55 | -10 | Frontal Lobe | Middle Frontal Gyrus | BA11 |
| 33 | -25 | 60 | 20 | Frontal Lobe | Superior Frontal Gyrus | BA10 |
| 34 | 30 | 60 | 15 | Frontal Lobe | Middle Frontal Gyrus | BA10 |
| 35 | 45 | 55 | -5 | Frontal Lobe | Middle Frontal Gyrus | BA10 |
| 36 | -55 | 35 | 5 | Frontal Lobe | Inferior Frontal Gyrus | BA45 |
| 37 | -20 | 45 | 45 | Frontal Lobe | Superior Frontal Gyrus | BA8 |
| 38 | 15 | 45 | 50 | Frontal Lobe | Superior Frontal Gyrus | BA8 |
| 39 | 50 | 45 | 10 | Frontal Lobe | Middle Frontal Gyrus | BA46 |
| 40 | -60 | 0 | -15 | Temporal Lobe | Middle Temporal Gyrus | BA21 |
| 41 | -45 | 15 | 50 | Frontal Lobe | Middle Frontal Gyrus | BA8 |
| 42 | 50 | 15 | 45 | Frontal Lobe | Middle Frontal Gyrus | BA8 |
| 43 | 60 | 5 | -15 | Temporal Lobe | Middle Temporal Gyrus | BA21 |
| 44 | -65 | -15 | 30 | Parietal Lobe | Postcentral Gyrus | BA3 |
| 45 | -30 | -15 | 70 | Frontal Lobe | Precentral Gyrus | BA6 |
| 46 | 35 | -15 | 70 | Frontal Lobe | Precentral Gyrus | BA6 |
| 47 | 65 | -15 | 30 | Parietal Lobe | Postcentral Gyrus | BA3 |
| 48 | -65 | -45 | -10 | Temporal Lobe | Middle Temporal Gyrus | BA21 |
| 49 | -50 | -50 | 55 | Parietal Lobe | Inferior Parietal Lobule | BA40 |
| 50 | -5 | -50 | 70 | Parietal Lobe | Postcentral Gyrus | BA5 |
| 51 | 50 | -50 | 55 | Parietal Lobe | Inferior Parietal Lobule | BA40 |
| 52 | 70 | -35 | -10 | Temporal Lobe | Middle Temporal Gyrus | BA21 |
| 53 | -55 | -70 | 25 | Temporal Lobe | Middle Temporal Gyrus | BA39 |
| 54 | -25 | -65 | 65 | Parietal Lobe | Superior Parietal Lobule | BA7 |
| 55 | 25 | -65 | 65 | Parietal Lobe | Superior Parietal Lobule | BA7 |
| 56 | 50 | -75 | 30 | Temporal Lobe | Angular Gyrus | BA39 |
| 57 | -40 | -90 | 5 | Occipital Lobe | Middle Occipital Gyrus | BA19 |
| 58 | -30 | -90 | 30 | Occipital Lobe | Cuneus | BA19 |
| 59 | -5 | -90 | 35 | Occipital Lobe | Cuneus | BA19 |
| 60 | 35 | -85 | 35 | Parietal Lobe | Precuneus | BA19 |
| 61 | 40 | -90 | 5 | Occipital Lobe | Middle Occipital Gyrus | BA19 |

## 15. Acknowledgements

We are indebted to Dr Tomas Ros for his time and effort in the format conversion of the open access EEG recordings from Babayan et al (2019), into EDF format. This allowed a final conversion from EDF to human readable text format by using the software EDFbrowser (van Beelen 2022), thus making it accessible to analysis with the LORETA-KEY free academic software (https://www.uzh.ch/keyinst/).